\documentstyle[12pt,twoside]{article}
\input{amssym.def}
\input{amssym}
\input{psfig}

\newtheorem{theorem}{Theorem}
\newtheorem{lemma}{Lemma}
\newtheorem{proposition}{Proposition}
\newtheorem{corollary}{Corollary}

\newcommand{\R} {{\Bbb R}}
\newcommand{\Q} {{\Bbb Q}}
\newcommand{\Z} {{\Bbb Z}}
\newcommand{\N} {{\Bbb N}}
\newcommand{\mod}[1] {\ ({\rm mod}\ {#1})}
\newcommand{\mtwo} {\ ({\rm mod}\ 2)}
\newcommand{\bi} {billiard}
\newcommand{\me} {measure}
\newcommand{\tr} {trajector}
\newcommand{\eo} {escape orbit}
\newcommand{\eps} {\varepsilon}
\newcommand{\fat} {\phi_{\alpha}^{t}}
\newcommand{\ft} {\phi^{t}}
\newcommand{\skippar} {\par\medskip}
\newcommand{\qed} {\hfill{\sc Q.E.D.}}

\newcommand{\La}{\Omega}
\newcommand{\Ln}{\Omega^{(n)}}

\renewcommand{\o} {orbit}
\renewcommand{\iff} {if, and only if,\ }

\begin{document}

\title{
	An Infinite Step Billiard 
}
\author{
	Mirko Degli Esposti $^{a)}$, Gianluigi Del Magno $^{b)}$, Marco Lenci 
	$^{a,c)}$ \\ 
	\\ 
	$^{a)}$ Dipartimento di Matematica \\ 
	Universit\`a di Bologna \\ 
	40127 Bologna, ITALY \\ 
	\\ 
	$^{b)}$ Mathematics Department \\ 
	Georgia Institute of Technology \\ 
	Atlanta, GA \ 30332, U.S.A. \\ 
	\\ 
	$^{c)}$ Mathematics Department \\ 
	Princeton University \\ 
	Princeton, NJ \ 08544, U.S.A. \\
	\\ 
	{\footnotesize\tt E-mail: desposti@dm.unibo.it, magno@math.gatech.edu, 
	marco@math.princeton.edu} \\
}
\date{August 1997}

\maketitle

\begin{abstract}
	A class of non-compact \bi s is introduced, namely the {\em infinite 
	step \bi s}, i.e., systems of a point particle moving freely in the 
	domain $\Omega = \bigcup_{n\in\N} [n,n+1] \times [0,p_{n}]$, with 
	elastic reflections on the boundary; here $p_{0}=1,\ p_{n}>0$ and 
	$p_{n} \searrow 0$.     
	\par 
	After describing some generic ergodic features of these dynamical 
	systems, we turn to a more detailed study of the example $p_{n} = 
	2^{-n}$.  What plays an important role in this case are the so 
	called {\em \eo s}, that is, \o s going to $+\infty$ monotonically 
	in the $X$-velocity.  A fairly complete description of them is 
	given.  This enables us to prove some results concerning the 
	topology of the dynamics on the \bi.
\end{abstract}

\section{Introduction} \label{sec-intro}

\begin{figure}[ht]
	\centerline{\psfig{file=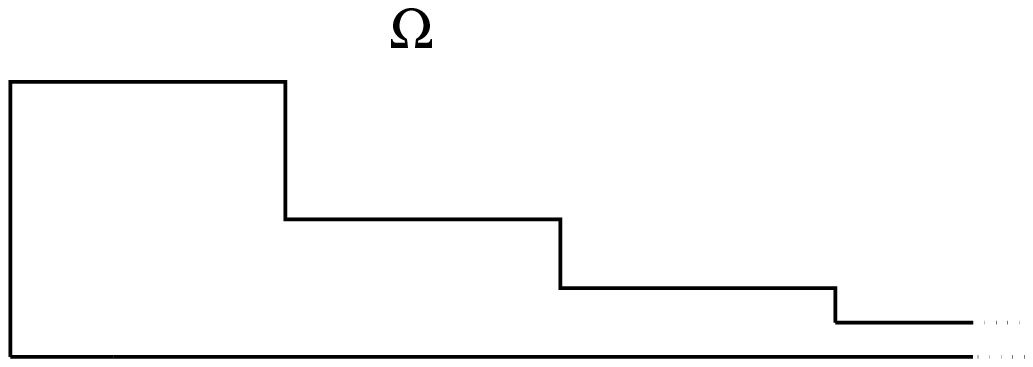}}
	\caption{The infinite billiard table $\Omega$.}
	\protect\label{fig1}
\end{figure}

Billiards are dynamical systems defined by the uniform motion of a 
point inside a domain with elastic reflections at the boundary, such 
that the tangential component of the velocity remains constant and the 
normal component changes sign.  The aim of this paper is to discuss 
some topological properties for a certain class of non-compact, 
polygonal \bi s, like the one depicted in Fig.  \ref{fig1}.
\par
Our main motivations originate from semiclassical quantum mechanics:
for example, it would be interesting to compare classical and quantum
localization for simple models of non-compact systems.  More
ambitiously, one might work in the direction of the semiclassical
asymptotics for the spectrum of the Hamiltonian operator: the
Gutzwiller trace formula and other semiclassical expansions \cite{gu}
relate this to the distribution of periodic \o s in the classical
system.  In the case of systems with {\em cusps}, similar to the ones
with which we are concerned here, these types of approximations become
more complicated, and one hopes to get a better understanding from the
knowledge of the \tr ies falling into the cusp ({\em \eo s}; see
\cite{le} and references therein).
\par
Finally, we believe that the investigation of the dynamical properties
of such kinds of models inherits an intrinsic interest by itself.
\par
In the case of a bounded polygonal \bi\ with a finite number of sites,
the \bi\ flow can be studied with the help of some well developed and
non-trivial techniques.  We refer to \cite{g2,g3} for the basic
definitions and results, reducing here to a brief and incomplete
review of some of them. Usually one assumes that the magnitude of the
particle's velocity equals one, and that the \o\ which hits a vertex
stops there (for our model, we will slightly modify this last
assumption).  However, the set of initial conditions whose \o s are
defined for all values of $t$, always represents a set of full \me\ in
the phase space.
\par
Among the class of polygonal \bi s, a \bi\ table $\Omega$ is a {\em
rational} \bi\ if the angles between the sides of $\Omega$ are all of the
form $\pi n_{i}/m_{i}$, where $n_{i}$ and $m_{i}$ are arbitrary
integers.  In this case, any \o\ will have only a finite number of
different angles of reflections.  Referring to \cite{g3} for a nice
review of the subject, we just note here that this rational condition
implies a decomposition of the phase space in a family of
flow-invariant surfaces $R_{\theta},\ 0 \le \theta \le \pi/m,
\ m:={\em l.c.m. \,\{m_i\}}$, planar representations of which
are obtained by the usual {\em unfolding} procedure for the \o s (see,
e.g., \cite{fk,zk}).  Excluding the particular cases $\theta =
0,\pi/m$, it is well known that the \bi\ flow restricted to any of the
$R_\theta$ is essentially equivalent to a geodesic flow
$\phi_{\theta}^{t}$ on a closed oriented surface $S$, endowed with a
flat Riemannian metric with {\em conical} singularities.  The
topological type of the surface $S$ (tiled by $2m$ copies of $\Omega$),
i.e., its genus $g$, is determined by the geometry of the rational
polygon.  For example, if $\Omega$ is a simple polygon, then
\begin{equation}
	g(S)=1+\frac{m}{2}\sum_{i=1,n}\frac{n_i-1}{m_i}.
	\label{genus}
\end{equation}
With the use of this equivalence, a number of theorems regarding the
existence and the number of ergodic invariant \me s for the flow have
been proven (\cite{zk} and references).
\par
More refined results concerning the \bi\ flows can then be obtained by
exploring the analogies of these flows with the {\em interval exchange
transformations} (using the induced map on the boundary) on one hand,
and with {\em holomorphic quadratic differentials} on compact Riemann
surfaces, on the other.
\par
The deep connections between these three different subjects have been 
proven very useful in the understanding of polygonal \bi\ flows.  In 
particular, we can summarize some of the most important statements in 
the next proposition (see \cite{g3} and references therein).  
[Briefly, let us recall that an {\em almost integrable \bi} is a \bi\ 
whose table is a finite connected union of pieces belonging to a 
tiling of the plane by reflection, e.g, a rectangular tiling, or a
tiling by equilateral triangles, etc.]

\begin{proposition} \label{mainRational}
The following statements hold true:
\begin{itemize}
	\item[(i)] {\rm \cite{kms,ar}} The Lebesgue \me\ in a (finite) 
	rational polygon is the unique ergodic \me\ for the \bi\ flow, for 
	(Lebesgue-)almost all directions.  
	\item[(ii)] {\rm \cite{zk}} For all but countably many directions, 
	a rational polygonal \bi\ is {\em minimal} (i.e., all infinite 
	semi-\o s are dense).  
	\item[(iii)] {\rm \cite{g1,g2,b}} For almost integrable \bi s, 
	``minimal directions'' and ``ergodic directions'' coincide.  
	\item[(iv)] {\rm \cite{gk}} Let ${\cal{R}}_{n}$ be the space of 
	$n$-gons such that their sides are either horizontal or vertical, 
	parametrized by the length of the sides.  Then for any direction 
	$\theta$, $0<\theta <\pi/2$, there is a dense $G_{\delta}$ in 
	${\cal{R}}_{n}$, such that for each polygon of this set the 
	corresponding flow $\phi_{\theta}^{t}$ is {\em weakly mixing}.  
	\item[(v)] {\rm \cite{ka}} For any rational polygon and any direction 
	$\theta$, the billiard flow $\phi_{\theta}^{t}$ is not {\em mixing}.
\end{itemize}
\end{proposition}

Moreover, by {\em approximating} generic polygons by rational
ones, other important results can be proven (we still refer to
\cite{g3} for a more exhaustive review):

\begin{proposition} \label{mainBilliard}
The following statements hold true:
\begin{itemize}
	\item[(i)] {\rm \cite{zk}} The set of transitive polygons is a dense 
	$G_{\delta}$.  
	\item[(ii)] {\rm \cite{kms}} For every $n$, there is a 
	dense $G_\delta$ of ergodic polygons with $n$ vertices.  
	\item[(iii)] {\rm \cite{g3}} For any given polygon, the metric 
	entropy with respect to any flow-invariant measure is zero.  
	\item[(iv)] {\rm \cite{gkt}} Given an arbitrary polygon and an \o, 
	either the \o\ is periodic or its closure contains at least one vertex.
\end{itemize}
\end{proposition}

In this paper we are interested in a class of rational \bi s, the {\em
infinite step \bi s}, defined as follows: let $\{p_n\}_{n\in\N}$ be a
monotonically vanishing sequence of positive numbers, with $p_{0}=1$.
We denote $\La := \bigcup_{n\in\N} [n,n+1] \times [0,p_{n}]$
(Fig. \ref{fig1}) and we call $(X,Y)$ the two coordinates on it.
\par
Following all the above considerations, we see that a point particle
can travel within $\Omega$ only in four directions (two if the motion
is vertical or horizontal---cases which we disregard).  One of these
directions lies in the first quadrant.  Therefore, for $\theta\in
]0,\pi/2[$ and $\alpha=\tan\theta$, the invariant surface is labeled
by $R_{\alpha}$ and is built via the unfolding procedure with four
copies of $\Omega$.  It can be represented on a plane $(X,Y)$ as
in Fig.\ref{fig2}, with the proper side identifications, and the $3\pi/2$
corners represent the {\em non-removable singularities}. With the 
additional condition $\sum_n p_n < \infty$, $R_{\alpha}$ can be considered
a non-compact, finite-area surface of infinite genus.
\par
We will denote by $\Ln$ the truncated \bi\ that one obtains
by closing the table at $X=n$.  The corresponding invariant surface
will be obviously denoted by $R_{\alpha}^{(n)}$ (Fig. \ref{fig3}) and
(\ref{genus}) shows that it has genus $n$.  Only to $R_{\alpha}^{(n)}$
can we apply the many strong statements of Proposition
\ref{mainRational} (see also Proposition \ref{prop-truncated} below).
Hence our interest in trying to extend some of those results to
the non-compact case. This paper gives a contribution in this
direction.
\par
After showing that examples can be given of infinite \bi s with the
above ergodic properties, we turn to the study of a billiard with
exponentially decreasing rational heights ($p_{n} = 2^{-n}$) and we give
some description of the topological behavior of its \o s.  More
precisely, we will first describe the existence and the number of the
so-called \eo s, showing that generically (in the initial directions)
there is exactly one \tr y ``traveling directly to infinity'' (Theorem
\ref{thm-generic}). This result makes use, among other more specific
computations, of a suitable family of interval maps ({\em rescaled 
transfer maps}), related to the return map to the first vertical wall.  
With the same tools, we then obtain a characterization of the behavior 
in the past for these unique \eo s (Theorem \ref{thm-backwards}).  
Finally, we analyze some topological properties for the flow associated 
to the infinite \bi.  The main outcome concerning this part is that 
the dynamics of the whole system is driven by the \eo\ which turns out 
to be a topologically complex object (Theorem \ref{thm-attractor}).

\subsection{General results} \label{subs-general}

Concerning the truncated billiards, we can put together some
of the previous results to state the following:

\begin{proposition}
	Fix $n\in\N$ and suppose $p_{k} \in\Q, \, \forall k\le n$.
	Consider the \bi\ $\Omega^{(n)}$.  If $\alpha \in \Q$, all
	the \tr ies are periodic.  If $\alpha \not\in \Q$, the flow is
	minimal and the Lebesgue \me\ is the unique invariant ergodic \me.
\label{prop-truncated}
\end{proposition}

{\sc Proof.} As already outlined, this proposition can be derived from
quite a number of results in the literature.  However, to give an
exact reference, \cite{g1}, Theorem 3 contains the assertion, since
$\Omega^{(n)}$ is an {\em almost integrable} \bi\ table. 
\par
It may be interesting to remark that the ideas on which the proofs are
based were already known sixty years ago, as \cite{fk} witnesses.  The
invariant surface $R_{\alpha}^{(n)}$ is divided into a finite number
of strips, that are either minimal sets or collections of periodic \o s 
(the two cases cannot occur simultaneously for an almost integrable
\bi).  These strips are delimited by {\em generalized diagonals}, that
is, pieces of \tr y that connect two (possibly coincident) singular
vertices of the invariant surface.  The above is nowadays called {\em the
structure theorem} for rational \bi s, a sharp formulation of which is
found, e.g., in \cite{ag}.
\par
Using this, minimality is easily established when, for a given
direction, no generalized diagonals and no periodic \o s are found.
\qed
\skippar
The above proposition will be used repeatedly during the remainder,
being more or less the only result we can borrow from our (much
wider) knowledge of the compact case. One of the first statements
we can derive from it is that we can actually find examples of step
\bi s which enjoy the ergodic properties one would expect. The price
we pay is that we must let the system decide, for a given irrational
direction, how fast the $p_{n}$'s should decay.

\begin{theorem}
	Fix $\alpha\not\in\Q$.  For every positive vanishing sequence
	$\{ \bar{p}_{n} \}$, there exists a strictly decreasing sequence
	$\{ p_{n} \} \subset \Q$, with $0 < p_{n} \le \bar{p}_{n}$, such that
	the \bi\ flow $\fat$ on $R_{\alpha}$, constructed as above according
	to $\{ p_{n} \}$, is ergodic (hence almost all \o s are dense).
	\label{thm-fast-decay}
\end{theorem}

The proof of this theorem is postponed to the next section, after we
have established some further notation.
\par
Another useful result can be derived from Proposition \ref{prop-truncated}:

\begin{proposition}
	Let an infinite step \bi\ $\Omega$ with rational heights ($p_{n}
		\in\Q,\: \forall n$) be given. If $\alpha \in \Q$, a semi-\o\ can be
	either periodic or unbounded. If $\alpha \not\in \Q$, all semi-\o s are
	unbounded.
	\label{prop-unbounded}
\end{proposition}

{\sc Proof.} If $\alpha \in \Q$ and we had a non-periodic bounded \tr
y, this would naturally correspond to a \tr y of $R_{\alpha}^{(n)}$,
for some $n\in\N$, which has only periodic \o s.  On the other hand,
if $\alpha \not\in \Q$, the dynamics over each $R_{\alpha}^{(n)}$ is
minimal.  Hence, every semi-\tr y reaches the abscissa $X=n$.
\qed

\subsection{The Return Map} \label{subs-rm}

In our realization of the surface $R_{\alpha}$, the first vertical
side of $\Omega$ becomes the closed curve $L :=\{0\} \times [-1,1[$
[$(0,-1)$ e $(0,1)$ are identified in Fig. \ref{fig2}] which
separates $R_{\alpha}$ in two symmetric parts. We will occasionally
identify $L$ with the interval $[-1,1[$.
\par
Except for the trivial case $\alpha = +\infty$ (vertical 
\o s---already excluded at the beginning), every \tr y crosses $L$ at
least once.  Without loss of generality, we will always assume to have
an initial point $(0,Y_{0})$ on the leftmost wall $L$, uniquely
associated to a pair $(Y_{0},\alpha) \in [-1,1[ \times ]0,+\infty[$.
We then use the Lebesgue \me\ as a natural way to \me\ \o s.
\par
The \bi\ flow along a direction $\alpha$, which we denote by $\fat$
(or $\ft$ when there is no means of confusion), induces a.e. on $L$ a
Poincar\'e map $P_{\alpha}$ that preserves the Lebesgue \me.  We call
it the {\em (first) return map}.  This discontinuous map is easily
seen to be an infinite partition {\em interval exchange
transformation} (i.e.t.).  On $L$ we establish the convention that the
map is continuous from above:  i.e., an \o\ going to the singular
vertex $(n,p_{n})$ of $R_{\alpha}$ will continue from the point
$(-n,p_{n})$, thus behaving like the \o s above it, i.e., bouncing
backwards. In the same spirit, a \tr y hitting $(-n,p_{n})$ will
continue from $(-n,-p_{n})$, while \o s encountering vertex
$(n,-p_{n})$ will just pass through. This corresponds to partitioning
$L \simeq [-1,1[$ into right-open subintervals.
\par
The fact that the number of subintervals is infinite is exactly
what makes the study of the ergodic properties of this system a
non-trivial task.
\par
It is now natural to relate $P_{\alpha}$ to the family of return maps
$P_{\alpha}^{(n)}$ corresponding to the truncated \bi s
$\Omega^{(n)}$.  These are finite partition i.e.t.'s defined on all of
$L$ (with abuse of notation, $L$ also denotes the obvious closed curve
on $R_{\alpha}^{(n)}$, Fig. \ref{fig3}).
\par
Let $E_{\alpha}^{(n)} \subset L$ be the set of points whose forward \o\ 
starts along the direction $\alpha$ and reaches the $n$-th {\em aperture}
$G_{n} := \{ n \} \times [-p_{n},p_{n}[$ without colliding with any
vertical walls.  $E_{\alpha}^{(n)}$ is union of at most $n$ right-open
intervals, since the backward evolution of $G_{n}$ can only {\em
split} once for each of the $n-1$ singular vertices (Fig.
\ref{fig4}).  We denote this by $n.i.(E_{\alpha}^{(n)}) \le n$, where
$n.i.$ stands for ``number of intervals''.  Moreover,
$|E_{\alpha}^{(n)}| = 2p_{n}$ and $E_{\alpha}^{(n+1)} \subset
E_{\alpha}^{(n)}$.  From this we infer that the family $\{
E_{\alpha}^{(n)} \}_{n>0}$ can be rearranged into sequences of nested
right-open intervals, whose lengths vanish as $n \to\infty$.
Clearly, the sequence of i.e.t.'s $P_{\alpha}^{(n)}\to P_{\alpha}$ a.e. in
$L$ as $n\to\infty$.
\skippar
The subset of $L$ on which $P_{\alpha}$ is not defined will be denoted
by $E_{\alpha} := \bigcap_{n>0} E_{\alpha}^{(n)}$ and clearly
$|E_{\alpha}|=0$ .  Each point of this set is the limit of an infinite
sequence of nested vanishing right-open intervals (the constituents of
the sets $E_{\alpha}^{(n)}$).  Elementary topology arguments allow us to
assert an almost converse statement: each infinite sequence yields a
point of $E_{\alpha}$, unless the ``pathological'' property holds that
the intervals eventually share their right extremes.
\par
The \o s starting from such points will never collide with
any vertical side of $R_{\alpha}$ (or $\Omega$) and thus, as $t\to
+\infty$, will go to infinity, maintaining a positive constant
$X$-velocity.  We call them {\em \eo s}.
\par
We now give the proof of Theorem 1.

\subsection{Proof of Theorem 1} \label{subs-proof-of-1}

We will construct $R_{\alpha}$ in such a way that almost every point
in $L$ has a typical \tr y, in the sense that the time average of a
function in a dense subspace of $L^{1}(R_{\alpha})$ equals its
spatial average.  Since $L$ is a Poincar\'e section, the same
property will hold for a.e.  point in $R_{\alpha}$.  For the sake of
notation, we will drop the subscript $\alpha$ in the sequel.
\par
Take a positive sequence $\eps_{n} \searrow 0$.  We are going to
build our \bi\ by induction: suppose we have fixed $p_{i}$ for $1
\le i \le n$, and we have to determine a suitable $p_{n+1}$. Consider
$R^{(n)}$, generated by the $p_{i}$'s found so far. The flow
$\phi_{(n)}^{t}$ on it is ergodic by Proposition
\ref{prop-truncated}. For $f\in L^{1}(R^{(n)})$ and $z\in L$ define
\begin{equation}
	\left( \Xi_{(n)}^{T} f \right) (z) := \frac{1}{T} \int_{0}^{T}
	f \circ \phi_{(n)}^{t} (z) \,dt - \frac{1}{|R^{(n)}|}
    \int_{R^{(n)}} f \, dXdY.
    \label{def-xi}
\end{equation}
Let $\{ f_{j}^{(n)} \}_{j\in\N}$ be a separable basis of
$L^{1}(R^{(n)})$.  For the rest of the proof $R^{(n)}$ will be
liberally regarded as an (open) submanifold of $R^{(m)}, m>n$.  As a
consequence, a function defined on the former set will be implicitly
extended to the latter by setting it null on the difference set. With
this in mind, let
\begin{equation}
	A_{T}^{(n)} := \left\{ z\in L \,|\, \forall\, 1 \le i,j \le n,
	\left| \, \Xi_{(n)}^{T} f_{i}^{(j)} (z) \, \right| \le \eps_{n}
	\right\}.
	\label{a-tau-n}
\end{equation}
By ergodicity, since only a finite number of functions are involved
in the above set, we have $|A_{T}^{(n)}| \to |L|=2$ as $T
\to \infty$. Take $T_{n}$ such that $|A_{T_{n}}^{(n)}| \ge
2-\eps_{n}/2$. We are now in position to determine $p_{n+1}$. Choose
some
\begin{equation}
	p_{n+1} \in\Q; \ 0 < p_{n+1} \le \min \left\{ \bar{p}_{n+1},
	\frac{\eps_{n}}{2 T_{n}} \right\},
	\label{p-n+1}
\end{equation}
and imagine to open a hole of width $2 p_{n+1}$ in the middle of
$\{n+1\} \times [-p_{n},p_{n}[$ (same as $\{-n-1\} \times
[-p_{n},p_{n}[$ since they are identified at the moment).  The motion
on $R^{(n)}$ is not affected very much by this change, during the time 
$T_{n}$.  If we denote by $\phi^{t}$ the flow on the infinite \bi\
table (when we are done constructing it), we can already assert that, taken
a point $z\in L$, $\phi_{(n)}^{t}(z) = \phi^{t}(z) \, \forall t\in
[0,T_{n}]$ unless the particle departing form $z$ hits the hole in a
time less than $T_{n}$.  We can estimate the \me\ of these ``unlucky''
initial points: they constitute the set
\begin{equation}
	B_{n} := L \cap \left( \bigcup_{t\in [-T_{n},0]}
	\phi_{(n)}^{t} ( \{n+1\} \times [-p_{n+1},p_{n+1}[ \,) \right).
\end{equation}
The backward beam (up to time $-T_{n}$) originating from the hole
cannot hit $L$ more than $T_{n}/2$ times, since between each two
successive crossings of $L$, the beam has to cover a distance which is
at least 2 (see Fig. \ref{fig3}), but the velocity of the particles was
conventionally fixed to 1. Every intersection of the beam with $L$ is
a set of \me\ $2 p_{n+1}$, so, from (\ref{p-n+1}), $|B_{n}| \le
\eps_{n}/2$.
\par
Set $C_{n} := A_{T_{n}}^{(n)} \setminus B_{n}$, thus $|C_{n}| \ge
2-\eps_{n}$.  So $C_{n}$ is the set of points which keep enjoying the
properties as in (\ref{a-tau-n}), even after the cut has been done in
$R_{(n)}$.  Suppose one repeats the above recursive chain of
definitions for all $n$ in order to define the infinite manifold $R$.
Let $C := \bigcap_{n\in\N} \bigcup_{m\ge n} C_{m}$.  Then $|C|=
\lim_{n\to\infty} \cup_{m\ge n} C_{m} = 2 = |L|$.  $C$ may be
called the {\em event} $\{ \{ C_{n} \}$ infinitely often$\}$;
it is the ``good'' set since, fixed $z\in C$, there exist a
subsequence $\{ n_{k} \}$ such that $z\in \bigcap_{k} C_{n_{k}}$.
This means that, taken two integers $i,j$, $\forall n_{k} \ge
\max\{i,j\}$,
\begin{equation}
	\left| \frac{1}{T_{n_{k}}} \int_{0}^{T_{n_{k}}} f_{i}^{(j)} \circ
	\phi^{t} (z) \,dt - \frac{1}{|R^{(n_{k})}|} \int_{R} f_{i}^{(j)} \,
	dXdY \right| \le \eps_{n_{k}}.
	\label{almost-done}
\end{equation}
Comparing this with (\ref{def-xi}) we notice two differences. First,
the flow that appears here is $\phi^{t}$ because of the remark after
(\ref{p-n+1}). Second, the manifold integral is taken over all of
$R$: this is so because of the initial convention to extend with zero
all functions defined on submanifolds of $R$.
\par
Define $\Xi^{T}$ in analogy with (\ref{def-xi}).  Since $|R^{(n)}|
\nearrow |R|$, (\ref{almost-done}) shows that $(\Xi^{T_{n_{k}}}
f_{i}^{(j)}) (z) \to 0$, as $k\to\infty$, with $T_{n}$ in general
going to $\infty$ (this is not indeed guaranteed by the definition of
$T_{n}$, but one can easily arrange to make this happen).  We would
not be done yet, if it were not for Birkhoff's Theorem, which
states that, for the function $f_{i}^{(j)}\in L^{1}(R)$, the time
average is well-defined a.e. (in $R$, hence in $L$). Summarizing,
for every $f \in {\rm span} \{ f_{i}^{(j)} \}_{i,j\in\N}$, there exists
a set $C_{f} \subseteq L,\, |C_{f}|=2$ such that
\begin{equation}
	\lim_{T\to +\infty} (\Xi^{T} f)(z) =0.
\end{equation}
This proves the claim we made in the beginning.  Since $\Xi^{T}$ is a
continuous operator in $L^{1}$ and ${\rm span} \{ f_{i}^{(j)} \}$ is
dense in it, we obtain the ergodicity part in the statement of Theorem
\ref{thm-fast-decay}.  As concerns the density result, this
immediately follows from standard arguments as in \cite{w}, Theorem
5.15 (which can be checked to hold under our hypotheses, as well).
\qed
\skippar
{\sc Remark.} The fact that the above result provides ergodic \bi s
with rational heights only is merely technical. We decided to use
Proposition \ref{prop-truncated}, which only deals with almost
integrable \bi s. For a generic finite step \bi\ one can as well say that
for almost all directions the flow is ergodic, by \cite{kms}, and a
slight generalization of Theorem \ref{thm-fast-decay} can be proven
in a hardly different way.

\section{The Exponential Step Billiard} \label{sec-exp}

\begin{figure}[ht]
	\centerline{\psfig{file=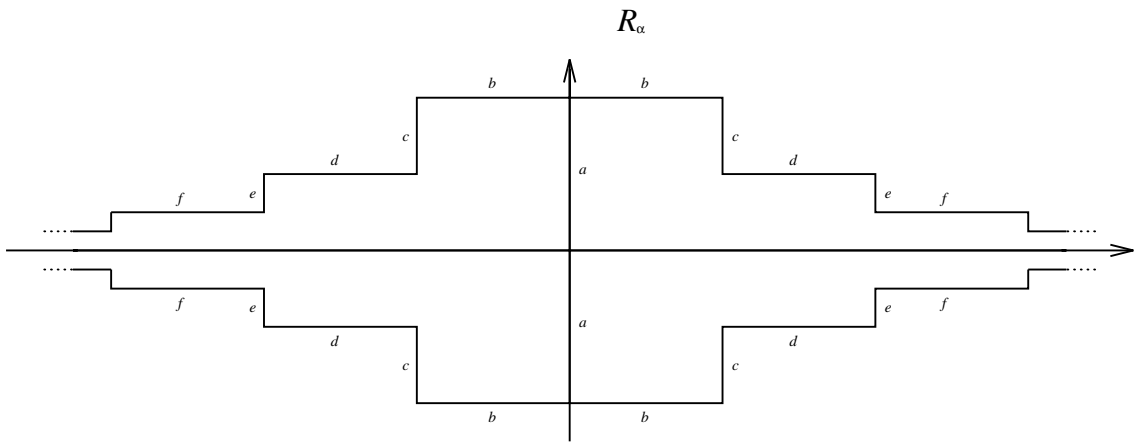}}
	\caption{The invariant surface $R_{\alpha}$ for the infinite billiard.}
	\protect\label{fig2}
\end{figure}

\begin{figure}[ht]
	\centerline{\psfig{file=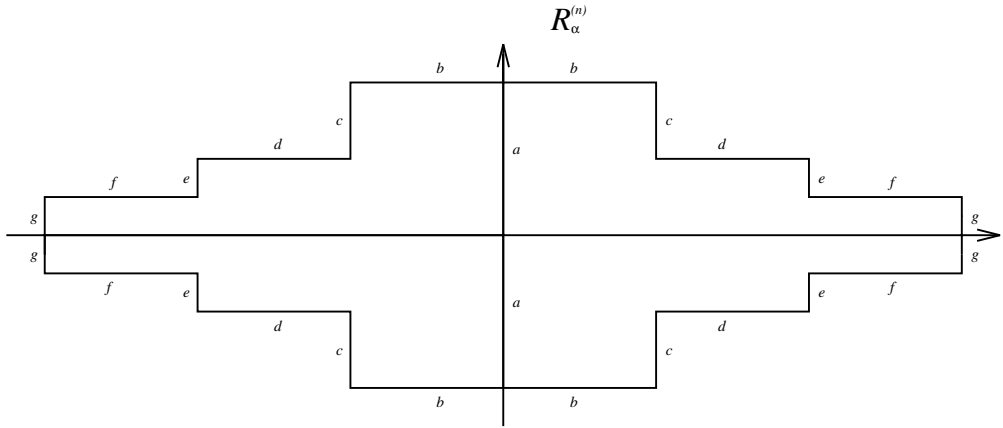}}
	\caption{The invariant surface $R_{\alpha}^{(n)}$ for the truncated 
	billiard.}
	\protect\label{fig3}
\end{figure}

\begin{figure}[ht]
	\centerline{\psfig{file=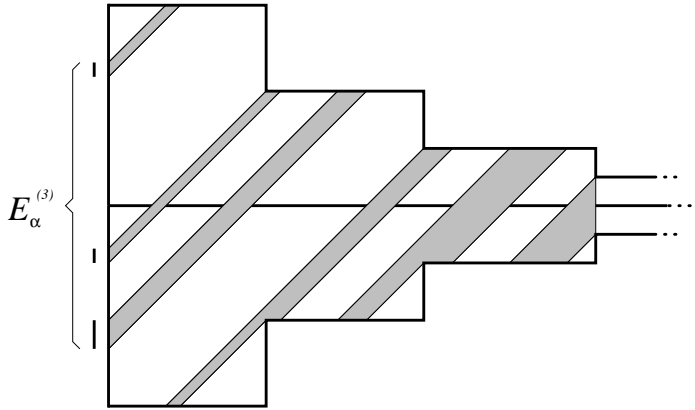}}
	\caption{Construction of $E_{\alpha}^{(n)}$ as the backward evolution 
	of the ``aperture'' $G_{n}$. The beam of orbits may split at 
	singular vertices.}
	\protect\label{fig4}
\end{figure}

Our main example of step \bi, to which we will restrict our attention
for the rest of this paper, is the {\em exponential step \bi}, i.e.,
the case $p_{n} = 2^{-n}$, shown in Figs. \ref{fig1}, \ref{fig2},
\ref{fig3}.
\par
We are, of course, interested in getting information about the unbounded
\o s of our non-compact dynamical system over $\Omega$, since bounded \o s
correspond to a \bi\ $\Omega^{(n)}$, for some $n\in\N$.  Indeed,
Proposition \ref{prop-unbounded} applies here allowing one to conclude
that (for almost all $\alpha$'s) all but a countable set of initial
conditions on $L$ give rise to unbounded \o s, which come back to
$L$ infinitely often.
\par
We will first focus on the \eo s, as introduced in Section
\ref{subs-rm}.  Strictly speaking, we consider only the asymptotic
behavior of the forward semi-\o.  But a glance at Fig. \ref{fig2} at
once shows that the backward semi-\o\ having initial conditions
$(Y_{0},\alpha)$ is uniquely associated, by symmetry around the
origin, to the forward semi-\o\ of $(-Y_{0},\alpha)$.
\skippar
{\sc Remark.} The above assertion needs to be better stated: although
the manifold $R_{\alpha}$ is symmetric around the origin, the flow
defined on it is not {\em exactly} invariant for time-reversal, as
Fig.  \ref{fig2} seems to suggest.  This is due to our convention in
Section \ref{subs-rm} about the continuity from above for the flow.
The time-reversed motion on $R_{\alpha}$ is isomorphic to the motion
on a manifold like $R_{\alpha}$ with the opposite convention
(continuity from below).  Nevertheless, little changes since only
singular \o s (a null-\me\ set) are going to be affected by this
slight asymmetry.
\skippar
We will characterize the existence and the number of the \eo s and
we will show that, generically, only one of the two branches of an \o\
can escape (see Section \ref{subs-back}).  Moreover, we borrow some
notation from \cite{l} and call {\em oscillating} all unbounded
non-escape (semi-)\o s.
\par
For the moment, let us introduce the following construction:
suppose that a \tr y $\gamma$ on $R_{\alpha}$ reaches {\em
directly}, that is monotonically in the $X$-coordinate, the opening
$G_{n} = \{n\} \times [-2^{-n},2^{-n}[$. Let us denote with $Y_{n} \in
[-2^{-n},2^{-n}[$ the ordinate of the point at which $\gamma$ crosses
$G_{n}$. Within the box $]n,n+1[ \times [-2^{-n},2^{-n}[$, the motion
is a simple translation. Hence
\begin{equation}
	Y_{n+1} = Y_{n} + \alpha \mod{2^{-n+1}},
	\label{pre-tm}
\end{equation}
with$\mod{r}$ meaning the {\em unique point} in $[-r/2,r/2[$ representing
the class of equivalence in $\R / r\Z$, rather than the class of
equivalence itself. The \tr y $\gamma$ will cross $G_{n+1}$ \iff
\begin{equation}
	Y_{n+1} \in [-2^{-(n+1)},2^{-(n+1)}[.
\end{equation}
Setting $y_{n} := 2^{n-1} Y_{n}$, relation (\ref{pre-tm}) becomes
\begin{equation}
	y_{n+1} = 2 y_{n} + 2^{n} \alpha \mtwo,
	\label{y_n+1-recurs}
\end{equation}
and the \tr y will cross $G_{n+1}$ \iff
\begin{equation}
	y_{n+1} \in \left[ -\frac{1}{2}, \frac{1}{2} \right[.
	\label{r-crossing}
\end{equation}
The recursion relation (\ref{y_n+1-recurs}) can be easily proven by 
induction to yield
\begin{equation}
	y_{n+1} = 2^{n+1} y_{0} + (n+1) 2^{n} \alpha \mtwo,
	\label{y_n+1-exact}
\end{equation}
where the numbers $y_{k}\in [-1/2, 1/2[$ now represent the (rescaled) 
intersections of the trajectory with the vertical openings $G_{k}$.
\par
The transformation $T_{n,\alpha}: [-1/2,1/2[ \, \longrightarrow [-1,1[$,
\begin{equation}
	T_{n,\alpha} (y) := 2 y + 2^{n} \alpha \mtwo
	\label{r-tm}
\end{equation}
will be called the {\em rescaled transfer map}.

\subsection{Escape Orbits} \label{subs-eo} 

In this section we shall exploit the \eo s for the exponential step
\bi\ $\Omega$ defined above. Actually, we will give a rather complete 
description of what happens to the set $E_{\alpha}$, as defined in 
Section \ref{subs-rm}, for $\alpha >0$.  We already know that 
$|E_{\alpha}| = 0$.  Among more detailed results, we will show that 
$E_{\alpha}$ can only contain one or two points, the former case 
holding for almost all directions $\alpha$.
\par 
We start with some easy statements, using rescaled coordinates, unless 
otherwise specified.

\begin{lemma}
	If $\alpha = 2k$, $k\in\Z$, only one \eo\ exists and its initial
	condition is $y_{0}=0$.
	\label{lemma-alpha=0}
\end{lemma}

{\sc Proof.} $T_{n,2k} = T_{n,0}\ \forall n\in\N$. The sequence $\{
y_{n} \}$ is in this case given by $y_{n}= 2^{n} y_{0} \mtwo$. If
$y_{0}=0$, all $y_{n}$ are null and the corresponding \tr y escapes,
according to (\ref{r-crossing}). If $y_{0} \ne 0$, there exists
a $k$ such that $|y_{k}| > 1/2$.
\qed
\skippar
In Fig.  \ref{fig2}, designate by $V_{n}$ the point $(n,-2^{-n}),\ n 
\ge 0$.  For $n\ge 1$, this means that, on the planar representation 
of $R_{\alpha}$, $V_{n}$ is the one copy of the $n$-th singular vertex 
of $\Omega$, such that its future semi-\o\ is going ``to the right''.

\begin{corollary}
	If $\alpha = k\, 2^{-j}$, with $k$ odd, $j$ non-negative integer,
	only one \eo\ occurs. This \o\ intercepts $V_{j}$.
	\label{cor-ver}
\end{corollary}

{\sc Proof.} The portion of the manifold $R_{\alpha}$ at the right of
the $(j+1)$-th aperture looks like $R_{\alpha}$ itself, modulo a
scale factor equal to $2^{-(j+1)}$.  Furthermore in that region, and
subject to the above rescaling, the transfer map is equivalent to the
one we have seen in the previous lemma.  In fact, for $n\ge j+1$,
$T_{n,\alpha} = T_{n,0}$.  So, to the part of the \eo\ after
$G_{j+1}$, we can apply that lemma and conclude that the escape
\tr y is unique and $y_{j+1}=0$ holds.  Now, we know that $\alpha$ is
indeed equal to $(2k'+1) 2^{-j}$.  Inverting (\ref{y_n+1-recurs}) with
$y_{j+1}=0$, we get $y_{j} = -k'-1/2-p$, for some integer $p$. By
(\ref{r-crossing}) $y_{j} \in [-1/2,1/2[$. Hence $y_{j} = -1/2$, which
proves the second part of the lemma.
\qed
\skippar
In Lemma \ref{lemma-alpha=0} we have encountered the case in which
$\{ T_{n,\alpha} \}$ is a sequence of identical maps. Considering the
more general case of a periodic sequence of maps will yield a useful
tool to detect the presence of more than one \eo.
\par
Observe that, when $\alpha=2k/(2^m-1)$ with $k,m$ positive 
integers, one gets $2^m \alpha=\alpha \mtwo$.  In this case we have a 
periodic sequence of transfer maps of period $m$, that is, 
$T_{pm,\alpha}=T_{0,\alpha}$ for all integer $p>0$.  For such directions, 
then, one method for detecting \eo s may be the following:  
Let us set $M_{m,\alpha} := T_{m-1,\alpha} \circ\:\cdots\:
\circ T_{0,\alpha}$. As in (\ref{y_n+1-exact}) it turns out that
$M_{m,\alpha}(y) = 2^{m}y + m 2^{m-1} \alpha \mtwo$. Let us now
find the fixed points of this map. Consider a \tr y having one of these
points as initial datum. If it crosses all openings between $G_{1}$
and $G_{m}$, then the sequence of crossing points $y_{0}, \ldots,
y_{m-1}$ will be indefinitely repeated and the \tr y will escape.
\par
Let us apply this technique to the case $k=1$ and $m=2$, that is $\alpha=2/3$. 
Hence the fixed points of the map $M_{2,2/3}$ are the points 
$y \in [-1,1[$ such that $y = 4y + 8/3 + 2p,\: p\in\Z$. They are
\begin{equation}
	y^{(0)} = -\frac{8}{9};\ \ y^{(1)} = -\frac{2}{9};\ \ y^{(2)} =
	\frac{4}{9}.
\end{equation}
Since $|y^{(0)}| = |M_{2,2/3}\, y^{(0)}| > 1/2$, that solution has to
be discarded. Instead, $y^{(1)} =: y^{(1)}_{0}$ is accepted since
\begin{eqnarray}
	y^{(1)}_{1} &=& 2 y^{(1)}_{0} + \frac{2}{3} \mtwo = \frac{2}{9}
	\in \left[ -\frac{1}{2},\frac{1}{2} \right[  \nonumber  \\
	y^{(1)}_{2} &=& y^{(1)}_{0} \in \left[ -\frac{1}{2},\frac{1}{2}
	\right[.
\end{eqnarray}
It turns out that the same holds for $y^{(2)}$. 
\par
Thus, for $\alpha=2/3$ there are at least two \eo s whose initial 
conditions in the non-rescaled coordinates are $Y_0 = -4/9$ and $Y_0 = 
8/9$.  As \o s of $R_{\alpha}$ they are distinct, but is this still 
true if we consider the corresponding \o s in the {\bi} $\Omega$?

\begin{lemma}
	For $\alpha=2/3$, the two \eo s with initial conditions $Y_0 = -4/9$ 
	and $Y_0 = 8/9$ have distinct projections on $\Omega$.   	 
	\label{lemma-dist}
\end{lemma}

{\sc Proof.} Suppose that the two \eo s coincide in $\Omega$.  Then 
the backward part of the orbit, that starts at $Y=-4/9$, must get to 
$Y=8/9$, after several oscillations.  According to the fact that a 
backward semi-\o\ having initial condition $(Y_0,\alpha)$ is associated 
to the forward semi-\o\ $(-Y_0,\alpha)$ (see remark at the beginning of 
Section \ref{sec-exp}), the geometry of $R_{\alpha}$ implies that
\begin{equation}
	\frac{8}{9} = -\frac{4}{9} + \sum_{i=1}^j \left(
	\frac{2}{3} - \frac{m_i}{2^{q_i}} \right),
\end{equation}
where $m_i,q_i$ are non-negative integers and $j$ is the number of 
rectangular boxes visited by the backward semi-\o\ before reaching 
the point with coordinate $Y=8/9$.  [We remind that in each box the 
variation of the $Y$-coordinate is $\alpha \mod{2^{-q_i}}$.] 
Rearranging this formula we obtain
\begin{equation}
	\frac{4}{9}=\frac{2}{3}j-\frac{m}{2^q},  
\end{equation}
for some non-negative integers $m$ and $q$. For any choice of $m,q$ and
$j$, the two sides are distinct. This contradicts our initial assumption that
the two \o s coincide. 
\qed
\skippar
We will see later in Corollary \ref{cor-le-2} that, along any direction
$\alpha$, there are no more than two \eo s in $R_{\alpha}$. We can summarize
everything about the case $\alpha=2/3$ in the following assertion. 

\begin{proposition}
	Along the direction $\alpha=2/3$ there are two distinct \eo s.
	\label{prop-2-eos}
\end{proposition}

We now turn to the study of the generic case.  Recalling the reasoning 
outlined in Section \ref{subs-rm} about the splitting of the backward 
beams of orbits (see also Fig.  \ref{fig4}), it comes natural to think 
that at this point we need to analyze the forward \tr ies starting 
from the singular vertices $V_{p} = (p,-2^{-p})$.  We name them 
$\gamma_{p}$.
\par
\begin{figure}[ht]
	\centerline{\psfig{file=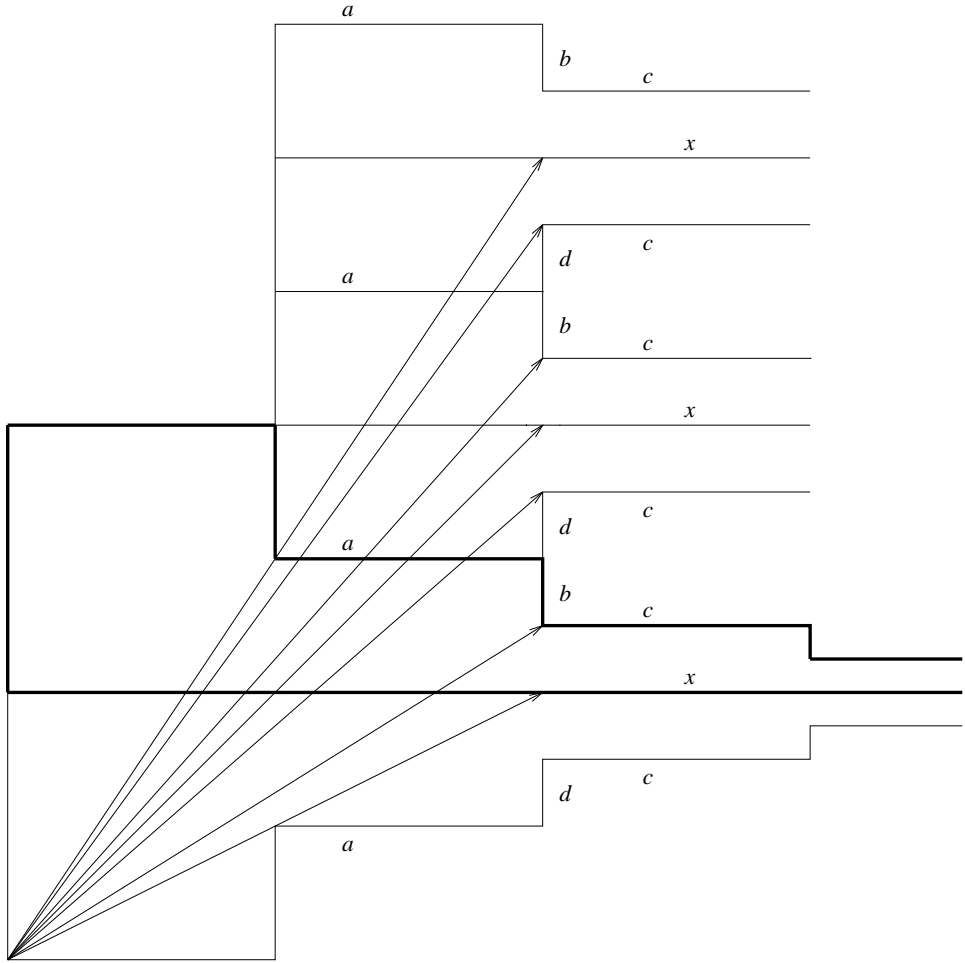}}
	\caption{Range of directions for which the orbit starting from
	the leftmost bottom vertex reaches directly aperture $G_{n}$
	(case $n=2$ is displayed). The billiard $\Omega$ has been unfolded
	on the plane, that is, many copies of it are sketched, in order
	to draw trajectories as straight lines. Considering
	$\alpha \mtwo$, for each $n \ge 2$, three beams occur. Fixing
	one beam relative to $G_{n}$, the geometry of the billiard
	implies that one, and only one, sub-beam will also reach $G_{n+1}$. 
	Eventually, for $n\to\ +\infty$, these three beams narrow down 
	to the values $\alpha = 1/2,1,3/2$, the last of which is rejected 
	for our convention on the continuation of singular orbits.}
	\protect\label{fig6}
\end{figure}

First of all, we consider those $\alpha$'s for which $\gamma_{0}$ 
reaches {\em directly} $G_{n}$, i.e., before hitting any vertical 
wall: we look at Fig.  \ref{fig6}, which displays the ``unfolding'' of 
$R_{\alpha}$, (where an \o\ over $R_{\alpha}$ is turned into a 
straight line).  A direct evaluation with a ruler, furnishes the 
answer, that runs as follows:
\par
\begin{eqnarray}
	& n=1) \ & \frac{1}{2} \le \alpha \mtwo < \frac{3}{2} 
	\label{gamma0-1} \nonumber \\ \\
	& n\ge 2) \ & \left\{ 
	\begin{array}{l} 
		\displaystyle{ \frac{1}{2} \le \alpha \mtwo < \frac{1}{2} + 
		\frac{1}{n 2^{n}}; } \\ \\
		\displaystyle{ 1 -\frac{1}{n 2^{n}} \le \alpha \mtwo < 1 + 
		\frac{1}{n 2^{n}}; } \\ \\
		\displaystyle{ \frac{3}{2}-\frac{1}{n 2^{n}} \le \alpha 
		\mtwo < \frac{3}{2}. } 
	\end{array} \right.  
	\label{gamma0-2}
\end{eqnarray}
Due to the self-similarity of our infinite \bi, we can write down the
analogous inequalities for every other vertex $V_{p},\, p\ge 1$ by
rescaling (\ref{gamma0-1}) and (\ref{gamma0-2}). Thus $\gamma_{p}$
crosses $G_{m},\ m>p$ \iff
\begin{eqnarray}
	& m=p+1) \ & \! \frac{1}{2^{p+1}} \le \alpha \mod{2^{-p+1}} < 
	\frac{3}{2^{p+1}} \label{gammap-1} \nonumber \\ \\
	& m\ge p+2) \ &  \! \left\{ 
	\begin{array}{l} 
		\displaystyle{ \frac{1}{2^{p+1}} \le \alpha \mod{2^{-p+1}} <
		\frac{1}{2^{p+1}} \!+\! \frac{1}{(m \!-\! p) 2^{m}}; } \\ \\
		\displaystyle{ \frac{1}{2^{p}} \!-\! \frac{1}{(m \!-\! p) 2^{m}} 
		\le \alpha \mod{2^{-p+1}} < \frac{1}{2^{p}} \!+\! 
		\frac{1}{(m \!-\! p) 2^{m}}; } 
		\\ \\ 
		\displaystyle{ \frac{3}{2^{p+1}} \!-\! \frac{1}{(m \!-\! p) 2^{m}} 
		\le \alpha \mod{2^{-p+1}} < \frac{3}{2^{p+1}}. } 
	\end{array} \right.  
	\label{gammap-2}
\end{eqnarray}
Working out these relations is essentially all we need to reach the 
goal we have set for ourselves at the beginning of this section.
From now on, when we say that an \o\ $\gamma_{p}$ reaches or crosses 
an aperture $G_{m}$, we will always mean {\em directly}.

\begin{lemma}
	If $\gamma_p$ is an {\eo} then it is the only \eo.
	\label{lemma-sing}
\end{lemma}

{\sc Proof.} It follows from (\ref{gammap-1})-(\ref{gammap-2}) that 
$\gamma_{p}$ is an \eo\ \iff $\alpha \in \{2^{-p},2^{-p-1}\} 
\mod{2^{-p+1}}$.  But for such $\alpha$'s, Corollary \ref{cor-ver} 
states that there is only one \eo.
\qed

\begin{lemma}
	Let $m,p$ be two non-negative integers with $m \ge p+2$. If $\gamma_{p}$ 
	crosses $G_{m}$, then either $\gamma_{p+1}$ does not reach $G_{p+2}$ or 
	it crosses $G_{m}$, as well.
	\label{lemma-p+1}
\end{lemma}

{\sc Proof.} It suffices to prove the statement for $p=0$ and the 
reader can easily get convinced that the actual result follows by a 
rescaling. Set
\begin{eqnarray}
	I_{m}^{(1)} & := & \bigcup_{j\in\N} \left[ \frac{1}{2}+2j, 
	\frac{1}{2} + \frac{1}{m2^m}+2j \right[,
	\nonumber  \\
	I_{m}^{(2)} & := & \bigcup_{j\in\N} \left[ 1 - \frac{1}{m2^m}+2j, 
	1 + \frac{1}{m2^m}+2j \right[,
	\\
	I_{m}^{(3)} & := & \bigcup_{j\in\N} \left[ \frac{3}{2} - 
	\frac{1}{m2^m} + 2j, \frac{3}{2}+2j \right[.
	\nonumber
\end{eqnarray}
From (\ref{gamma0-2}), $\gamma_{0}$ crosses $G_{m}$ \iff $\alpha \in 
I_{m} := I_{m}^{(1)} \cup I_{m}^{(2)} \cup I_{m}^{(3)}$.  If $\alpha 
\in I_{m}^{(2)}$ then $\gamma_{1}$ does not cross $G_{2}$.  In fact, 
(\ref{gammap-1}) states that $\gamma_{1}$ crosses $G_2$ \iff $\alpha 
\in B := \bigcup_{k \in\N} [1/4+k, 3/4+k[$.  So we have to prove that 
the sets $I_{m}^{(2)}$ and $B$ have empty intersection.  This is the 
case, because $I_{m}^{(2)}$ is made up of intervals of center 
$2j+1$ and radius $1/(m2^m)$, and $B$ is made up of intervals of 
center $1/2+k$ and radius $1/4$, so that
\begin{equation}
	{\rm dist}(I_{m}^{(2)},B) \ge \frac{1}{2} - \left( \frac{1}{m2^m} 
	+ \frac{1}{4} \right) > 0 \ \ \ {\rm for} \ m \ge 2.
\end{equation}
It remains to analyze the case $\alpha \in C := I_{m}^{(1)} \cup 
I_{m}^{(3)}$.  Relations (\ref{gammap-2}) tell us that $\gamma_{1}$ 
crosses $G_{m}$ \iff
\begin{eqnarray}
	\alpha \in D := \bigcup_{k\in\N} &\Biggl(& \left[ 
	\frac{1}{4} + k, \frac{1}{4} + \frac{1}{(m-1)2^m} + k \right[ \cup
	\nonumber  \\
	&\cup & \left[ \frac{1}{2} - \frac{1}{(m-1)2^m}+k, \frac{1}{2} + 
	\frac{1}{(m-1)2^m} + k \right[ \cup
	\label{def-D} \\
	&\cup& \left[ \frac{3}{4} - \frac{1}{m2^m} + k, \frac{3}{4} + k
	\right[ \ \ \Biggr).
	\nonumber
\end{eqnarray}
We have to prove that $C \subseteq D$.  We can visualize the sets $C$ 
and $D$ as periodic structures on the line whose fundamental patterns 
have, respectively, lengths 2 and 1 (with common endpoints).  Therefore, 
defining $\hat{C} := C \cap [0,2] = [1/2, 1/2 + 1/(m2^m)[ \, \cup \,
[3/2 - 1/(m2^m), 3/2[$ and $\hat{D} := D \cap [0,2]$, all we have to do 
is to show that $\hat{C} \subseteq \hat{D}$.  Deducing the shape of 
$\hat{D}$ from (\ref{def-D}), the result follows from the trivial 
relations:
\begin{eqnarray}
	\left[ \frac{1}{2}, \frac{1}{2} + \frac{1}{m2^m} \right[ & 
	\subset & \left[ \frac{1}{2} - \frac{1}{(m-1)2^m}, \frac{1}{2} + 
	\frac{1}{(m-1)2^m} \right[,
	\\
	\left[ \frac{3}{2} - \frac{1}{m2^m}, \frac{3}{2} \right[ & \subset & 
	\left[ \frac{3}{2} - \frac{1}{(m-1)2^m}, \frac{3}{2} + 
	\frac{1}{(m-1)2^m} \right[.
\end{eqnarray}
\qed

\begin{lemma}
	Again $m \ge p+2$. If $\gamma_{p}$ crosses $G_{m}$, then for all 
	$p+2\le n\le m$, $\gamma_{n}$ does not reach $G_{n+1}$.
\label{lemma-geo}
\end{lemma}

{\sc Proof.} As before, we give the proof only for the case $p=0$.  
The \o\ $\gamma_{0}$ crosses $G_m$ \iff $\alpha \in I_{m}$ defined in 
the proof of the previous lemma, whereas $\gamma_{n}$ crosses 
$G_{n+1}$ \iff
\begin{equation}
	\alpha \in J_{n} := \bigcup_{k\in\N} \left[ \frac{1}{2^{n+1}} +
	\frac{k}{2^{n-1}}, \frac{3}{2^{n+1}} + \frac{k}{2^{n-1}} \right[.
\end{equation}
If $I_{m}$ and $J_{n}$ have empty intersection, for all $2\le n \le 
m$, then the lemma is proven. Proceeding as in the first part of Lemma 
\ref{lemma-p+1}, we see that $I_{m}$ is strictly contained in a union of 
intervals of center $q/2$ and radius $1/(m2^m)$, while the intervals 
constituting $J_{n}$ have center $2^{-n}+ k 2^{-n+1}$ and radius 
$2^{-n-1}$. Thus, for $3\le n \le m$,
\begin{equation}
	{\rm dist}(I_{m},J_{n}) \ge \frac{1}{2^n} - \left( \frac{1}{m2^m} 
	+ \frac{1}{2^{n+1}} \right) > 0.
	\label{dist}
\end{equation}
If $n=2$, (\ref{dist}) becomes an equality, but the fact that our 
intervals are right-open ensures nevertheless that $I_{m} \cap J_{2}
= \emptyset$.
\qed
\skippar
One way to memorize the previous technical lemmas may be as follows.  
The fact that $\gamma_{p}$ crosses $G_{m}$ influences all 
$\gamma_{n}$'s, for $n$ between $p+1$ and $m$: if $\gamma_{p+1}$ wants 
to ``take off'' (that is, reach some apertures), then it is forced to 
follow, and possibly pass, $\gamma_{p}$; while the $\gamma_{n}$'s 
with $n \ge p+2$ cannot even take off.
\par
We now enter the core of the arguments: recall the notation $n.i.$ to 
designate the number of disjoint intervals that constitute a set.

\begin{lemma}
	Fix $\alpha>0$. Either there exists an integer $q$ such that 
	$n.i.(E_{\alpha}^{(n)}) = 2$ for all $n \ge q$, or there is a 
	sequence $\{n_{j}\}$ such that $n.i.(E_{\alpha}^{(n_{j})}) = 1$.
\label{lemma-seq}
\end{lemma}

{\sc Proof.} The set of $\alpha$'s with the property that $n.i.  
(E_{\alpha}^{(n)})=2$ for $n \ge q$ is not empty.  In fact, by direct 
computation, it is easy to verify that for $\alpha = 2/3$ every 
$\gamma_{n}$ crosses $G_{n+1}$ but not $G_{n+2}$, so that $n.i.  
(E_{\alpha}^{(n)})=2$ for all $n > 0$.
\par
Now, suppose there exists a sequence $\{{m_j}\}$ with 
$n.i.(E_{\alpha}^{(m_{j})}) \ne 2$.  We can assume 
$n.i.(E_{\alpha}^{(m_{j})}) \ge 3$, otherwise, maybe passing to a 
subsequence, we would have $n.i.(E_{\alpha}^{(m_{j})})=1$ and we would 
be done.  If we fix an $m_{j}$ there are at least two singular \o s 
that cross $G_{m_{j}}$.  Let $0 < p_{j} \le m_{j}-2$ be the smallest 
integer such that $\gamma_p$ crosses $G_{m_{j}}$.  It follows from 
Lemma \ref{lemma-geo} that only $\gamma_{p_{j}}$ and 
$\gamma_{p_{j}+1}$ cross $G_{m_{j}}$.  Therefore 
$n.i.(E_{\alpha}^{(m_{j})}) = 3$.
\par
At this point we have three cases: $\gamma_{p_{j}}$ and 
$\gamma_{p_{j}+1}$ are both \eo s; one of them escapes and the other 
is reflected; they are both reflected.
\par
In the first case Lemma \ref{lemma-sing} ensures that $\gamma_{p_{j}}$ 
and $\gamma_{p_{j}+1}$ coincide and Lemma \ref{lemma-geo} implies that 
no $\gamma_{n}$ with $n > p_{j}+1$ can ``take off''.  Hence 
$n.i.(E_{\alpha}^{(m)})=2$ for all $n\ge p_{j}+1$, contradicting our 
assumption. The second case is hardly any different: call $G_{n_{j}}$ 
the first aperture that $\gamma_{p_{j}}$ cannot reach (in fact Lemma 
\ref{lemma-p+1} implies that, of the two, $\gamma_{p_{j}+1}$ must be the 
escaping \tr y).  Therefore, using again Lemma \ref{lemma-geo}, 
$n.i.(E_{\alpha}^{(m)})=2$ for all $n\ge n_{j}$, a contradiction as 
before.  In the last case, call $G_{n_{j}}$ the first aperture which 
is not reached by $\gamma_{p_{j}+1}$, and thus not even by 
$\gamma_{p_{j}}$.  [Lemma \ref{lemma-p+1} claims that 
$\gamma_{p_{j}+1}$ goes farther than $\gamma_{p_{j}}$.] Another 
application of Lemma \ref{lemma-geo} proves that 
$n.i.(E_{\alpha}^{(n_{j})})=1$.  Proceeding inductively we find a 
sequence of integers $n_{j} > m_{j}$ with the desired property.
\qed

\begin{corollary}
	For all $\alpha$'s, $\# E_{\alpha} \le 2$.
	\label{cor-le-2}
\end{corollary}

\begin{lemma}
	Notation as in the above lemma. In the case $n.i.(E_{\alpha}^{(n)}) 
	= 2$ for $n \ge q$, suppose $q\ge 1$ is the minimum integer 
	enjoying that property. Then there are only two possibilities:
	\begin{itemize}
		\item[(a)]  $\gamma_{q-1}$ is the only \eo\ and $\alpha = 
		2^{1-q} \mod{2^{2-q}}$.
		\item[(b)]  $\gamma_{n}$ crosses $G_{n+1}$ but not $G_{n+2}$ 
		for all $n \ge q-1$ so that there are two \eo s and either 
		$\alpha = 2^{2-q}/3 \mod{2^{2-q}}$ or $\alpha = 2^{3-q}/3 
		\mod{2^{2-q}}$.
	\end{itemize}
	\label{lemma-ni-2}
\end{lemma}

{\sc Proof.} First, let us see that $\gamma_{q-1}$ is the only singular 
\o\ crossing $G_{q}$. In fact $G_{q}$, by hypothesis, is intersected 
by only one $\gamma_{k}$ ($k \le q-1$). [Actually, the case may occur 
that both $\gamma_{p}$ and $\gamma_{k}$ cross that aperture, but
only if they coincide. Nothing changes in the argument if we take 
$k$ to be the largest integer of the two.] If $k \le q-2$, then by 
Lemma \ref{lemma-geo}, no singular \o\ $\gamma_{n}$, with $n\ge k+2$ 
can ``take off''.  Neither can $\gamma_{k+1}$, which would be forced, 
by Lemma \ref{lemma-p+1}, to pass $G_{q}$, against the hypotheses.
The net result is that $n.i.(E_{\alpha}^{(n)}) =2,\ \forall n\ge 
k+1$, which contradicts the minimality of $q$.
\par
Now suppose that $\gamma_{q-1}$ reaches $G_{q+1}$.  We want to prove 
that it is also an \eo\ and we are in case {\em (a)}.  In fact, if it 
stops somewhere after $G_{q+1}$ (say right before $G_{k},\ k>q+1$), 
then $\gamma_{q}$ either passes it (and $n.i.(E_{\alpha}^{(q+1)}) = 
3$) or $\gamma_{q}$ does not ``take off'' (and 
$n.i.(E_{\alpha}^{(k)}) = 1$). Let us see for which directions this 
happens: from (\ref{gammap-2}), $\alpha\in \{ 2^{-q},2^{1-q} 
\} \mod{2^{2-q}}$; if $\alpha=2^{-q} \mod{2^{2-q}}$, then 
$n.i.(E_{\alpha}^{(q)})=1$, so that it must be $\alpha=2^{1-q} 
\mod{2^{2-q}}$.  Considering $\{E_{\alpha}^{(n)}\}_{n\in\N}$, it is 
easy to see that it consists of two nested sequences of right-open 
intervals.  One of the sequences collapses into the empty set, since all 
of the intervals share their right endpoint.
\par
So the remaining case is: $\gamma_{q-1}$ reaches $G_{q}$ but not
$G_{q+1}$. We would like to prove that this also occurs for all
$n>q-1$, i.e., we are in case {\em (b)}. With the same arguments as
above, one checks that either $\gamma_{q}$ reaches $G_{q+1}$, but not
$G_{q+2}$, or it escapes to $\infty$. The latter cannot be the case,
since we already know the only direction for which this can happen
[namely $\alpha=2^{-q} \mod{2^{1-q}}$]: this is the direction for which
$\gamma_{q-1}$ and $\gamma_{q}$ coincide, contrary to our present
assumption. Reasoning inductively, we obtain the assertion.
\par
Here, as before, we see that $\{E_{\alpha}^{(n)}\}_{n\in\N}$ consists 
of two nested sequences of right-open intervals.  But now the two 
intervals, for a given $n$, share alternatively [in $n$] the right and 
the left endpoint, so that each sequence shrinks to one point.  It 
remains to find the directions corresponding to this case.  In the 
sequel, without loss of generality, we assume $q=1$.
\par
\begin{figure}[ht]
	\centerline{\psfig{file=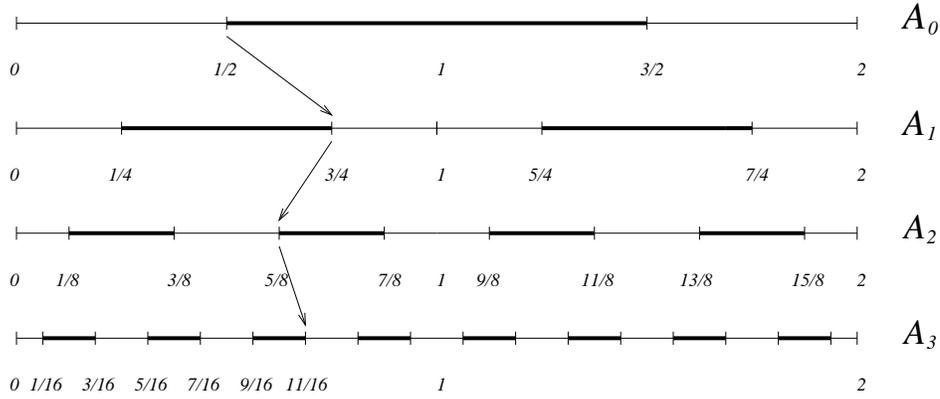}}
	\caption{The structure of the sets $A_{n}$, as in the proof of Lemma 
	\ref{lemma-ni-2}: $A=\bigcap_{n \ge 0} A_n$ consists of two points, 
	both limit of a sequence of nested intervals.}
	\protect\label{fig9}
\end{figure}

Let $A_n$ be the set of directions along which $\gamma_n$ crosses 
$G_{n+1}$, for all $n \ge 0$.  According to (\ref{gammap-1}), 
$A_n=\bigcup_{k\in\N} ([2^{-n-1},3 \, 2^{-n-1}[ + k \, 2^{1-n})$.  We 
claim that $A=\bigcap_{n \ge 0} A_n$ is the set of $\alpha$'s we are 
looking for.  In fact, if $\alpha \in A$ then $\gamma_n$ crosses 
$G_{n+1}$ for all $n \ge 0$, by definition of $A$.  Moreover 
$\gamma_n$ does not cross $G_{n+2}$, because if it did then, by Lemma
\ref{lemma-geo}, $\gamma_{n+2}$ would not cross $G_{n+3}$, which is a 
contradiction.  We note that every $A_n$ has a periodic structure 
whose fundamental pattern has length $2^{1-n}$. The least common 
multiple of these numbers is 2.  Thus, as in the proof of Lemma 
\ref{lemma-p+1}, we only need to look at $\hat{A} := A \cap 
[0,2]$.  This set consists of two points: $\alpha_1$ and $\alpha_2$.  
In fact, let $\hat{A}_{p} := [0,2] \cap (\bigcap_{n=0}^{p} A_n)$. 
Then, referring at Fig.  \ref{fig9}, it is clear that 
$\{\hat{A}_{p}\}$ is made of two sequences of nested intervals, both 
having a limit $\alpha_{i}$.  Furthermore, by the symmetry of the 
$A_n$'s, $\alpha_2 = 2-\alpha_1$.  As indicated by Fig.  \ref{fig9}, 
one way to find $\alpha_1$, and therefore $\alpha_2$, is to compute 
the limit of the oscillating sequence $2^{-1}, 2^{-1}+2^{-2}, 
2^{-1}+2^{-2}-2^{-3}, 2^{-1}+2^{-2}-2^{-3}+2^{-4}, \dots$  In other 
words, $\alpha_1=\sum_{j=0}^\infty 2^{-1-2j}=2/3$ so that 
$\alpha_2=4/3$.  Hence, for $q=1$, the directions that generate the 
behavior described in {\em (b)} are $\alpha=2/3 \mtwo$ and $\alpha=4/3
\mtwo$.
\qed
\skippar
This lemma has an important consequence which will be appreciated in 
Section \ref{sec-dyn}:

\begin{corollary}
	For almost every $\alpha$, one can find a sequence $\{ n_{j} \}$ 
	such that $n.i.(E_{\alpha}^{(n_{j})}) = 1$.
	\label{cor-main}
\end{corollary}

To complete the description of $\# E_{\alpha}$, we give now our last 
result:

\begin{proposition}
	There are no $\alpha$'s without \eo s.
	\label{prop-zero}
\end{proposition}

\begin{figure}[ht]
	\centerline{\psfig{file=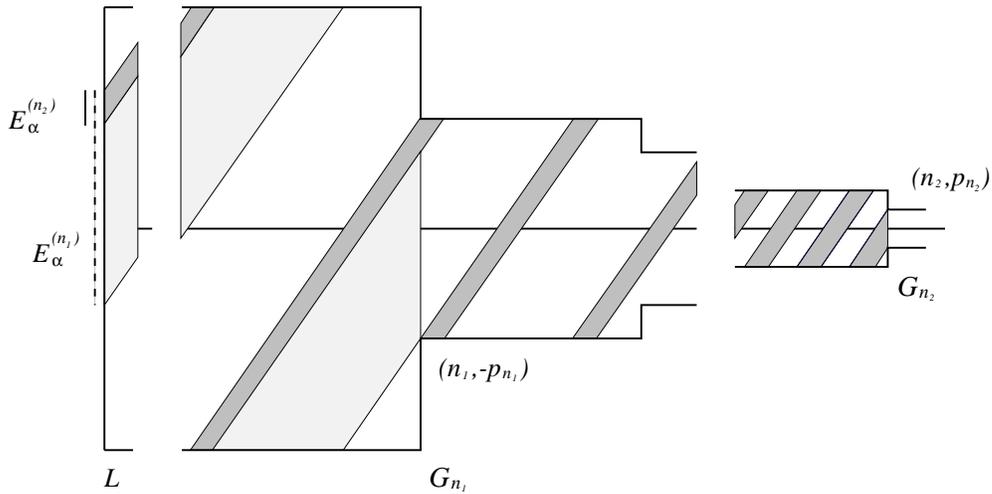}}
	\caption{In order to have no escape orbits, $E_{\alpha}^{(n_{1})}$, 
	$E_{\alpha}^{(n_{2})}$, etc. must have upper (equivalently right) 
	extremes in common. This implies that the beams of orbits departing 
	from them have upper boundaries in common, whence the existence of 
	pieces of generalized diagonal. An analysis of the directions 
	$\alpha$ for which this should happen shows that this is not the 
	case (Proposition \ref{prop-zero}).}
	\protect\label{fig5}
\end{figure}

{\sc Proof.} From the previous lemmas, there may be zero \eo s only 
for those $\alpha$'s such that there is a sequence $\{n_j\}$ with
$n.i.(E_{\alpha}^{(n_j)}) = 1$. Moreover, in order to have no \eo s,
the intervals $E_{\alpha}^{(n_j)}$ must eventually share their right 
extemes. This implies that $\gamma_{n_j}$ connects the vertex $V_{n_j}$ 
to the ``upper copy'' of $V_{n_{j+1}}$, as illustrated in Fig.
\ref{fig5}. Note that if $n_{j+1} - n_{j}=1$ for some $j \ge 0$, 
then $\gamma_j$ is an \eo\ (essentially the case {\em (a)} of Lemma 
\ref{lemma-ni-2}).  We can assume that $n_0=0$, otherwise can always 
rescale the \bi.  Thus $\gamma_0$ connects the vertices $V_{0}$ to 
$V_{n_1}$.  By looking at (\ref{gamma0-1})-(\ref{gamma0-2}), this 
happens \iff
\begin{equation}
	\alpha=\frac{1}{2} + \frac{1}{n_{1}2^{n_1}} + 2k_1 \ \ {\rm or} 
	\ \ \alpha=1+\frac{1}{n_{1}2^{n_1}} + 2k_1,
	\label{alpha-n1}
\end{equation}
for some integer $k_1$.  Now, let us consider $\gamma_1$.  If we 
rescale vertically the \bi\ by a factor $2^{n_1}$, we get the 
same setting we had for $\gamma_0$.  Since $\gamma_1 $connects 
$V_{n_1}$ to $V_{n_2}$, we must have, for some $k_{2}$,
\begin{equation}
	2^{n_1} \alpha = \frac{1}{2}+\frac{1}{(n_2-n_1)2^{n_2}}+2k_2 \ \ 
	{\rm or} \ \ 2^{n_1} \alpha = 1+\frac{1}{(n_2-n_1)2^{n_2}}+2k_2.
	\label{alpha-n2}
\end{equation}
Since $n_2-n_1 > 1$ and $n_1 > 1$, a comparison between 
(\ref{alpha-n1}) and (\ref{alpha-n2}) shows that $1/n_1$ must equal 
$1/2 + 1/((n_1-n_2) 2^{n_2})$ or $1 + 1/((n_1-n_2) 2^{n_2})$.  It is 
not hard to see that this cannot be the case.  Therefore there are no 
$\alpha$'s such that $\gamma_0$ intersects $V_{n_1}$ and $V_{n_2}$ at 
the same time.  This proves the statement.
\qed 
\skippar
Lemma \ref{lemma-seq}, Lemma \ref{lemma-ni-2} and Proposition
\ref{prop-zero}, can be summarized into the following theorem.

\begin{theorem}
	With reference to the step \bi\ $\Omega$: If there exists a
	non-negative integer $m$ such that $2^m \alpha = 4/3 \mtwo$, 
	then there are two \eo; otherwise there is only one \eo.  
	\label{thm-generic}
\end{theorem}

We osberve that, if there are two \eo s, they are distinct even in 
$\Omega$, by using the same considerations as in the proof of Lemma 
\ref{lemma-dist}. The core of the theorem, however, is re-expressed in 
the following important corollary.

\begin{corollary}
	For all but countably many $ \alpha > 0 $ the step billiard has
	exactly one \eo.
	\label{cor-gen}
\end{corollary}

\subsection{The backward part of an \eo} \label{subs-back}

In this section we explore the behavior of an \eo\ for negative times.
This question turns out to be crucial for the understanding of the
dynamics on the exponential step \bi, as it will explained in Section
\ref{sec-dyn}.
\par
As trivial as it is, we point out that the backward part of any \eo\
cannot be periodic; nor it can be bounded, by Proposition
\ref{prop-unbounded}. What about the possibility for it to escape to
$\infty$ as well, having a constant negative $X$-velocity for $t\in
]-\infty,t_{0}[$?

\begin{lemma}
	For a.e. $\alpha$, the backward part of an \eo\ does not intersect
	any vertex.
	\label{lemma-no-vertex}
\end{lemma}

{\sc Proof.} Fix an $\alpha$ for which the assertion does not hold: we 
then have in $R_{\alpha}$ an \eo\ containing a vertex $V$ before it 
``takes off'' towards infinity.  $V$ can only be $(0,0)$ or of the 
form $(p,\pm 2^{-q})$ (incidentally, Fig.  \ref{fig2} shows that 
$q=|p|$ or $q=|p|+1$).
\par
Let us call $(0,Y_{0})$ the last (in time) intersection point between
the \o\ and $L$.  The geometry of $R_{\alpha}$ implies that
\begin{equation}
	Y_{0} = m_{0} 2^{-q} + \sum_{i=1}^{j} (\alpha - m_{i} 2^{-q_{i}});
\end{equation}
with $m_{i},q_{i} \ge 0$ some integers ($m_{0}=0$ or $\pm 1$ depending 
on $V$ being the origin or not).  The above formula is easily 
understood thinking that $j$ is the number of rectangular boxes 
visited by the \o\ before reaching $L$ for the last time: in each box 
the variation of the $Y$-coordinate is $\alpha \mod{2^{-q_{i}}}$.  We 
turn now to the rescaled coordinates: $y_{0} = Y_{0}/2$.  Thus, 
rearranging the previous equality yields, for some integers $m,k$,
\begin{equation}
	y_{0} = m 2^{-k} + \frac{j}{2} \alpha.
	\label{y-zero}
\end{equation}
Since the \o\ is supposed to escape after leaving $(0,Y_{0})$, we can
apply (\ref{r-crossing}), (\ref{y_n+1-exact}) with $y_{0}$ as in
(\ref{y-zero}). If $n\ge k$ the first term in (\ref{y-zero}) gets canceled.
Therefore one must have
\begin{equation}
	y_{n+1} = (n+j+1) 2^{n} \alpha \mtwo \in \left[ -\frac{1}{2},
	\frac{1}{2} \right[ \ \ \ \forall n \ge k.
	\label{escape1}
\end{equation}
Consider the increasing sequence $\{ n_{i} \}_{i\ge p}$ such that
$n_{i}+j+1=2^{i}$ with $n_{p} \ge k$.  Condition (\ref{escape1}) implies
in particular that
\begin{equation}
	y_{n_{i}+1} = 2^{i+n_{i}} \alpha \mtwo \in \left[ -\frac{1}{2},
	\frac{1}{2} \right[ \ \ \ \forall i \ge p.
	\label{escape2}
\end{equation}
By looking at the appendix---especially at Lemma
\ref{lemma-expans}---one easily sees that (\ref{escape2}) is equivalent
to saying that the $-(i+n_{i}+1)$-th digit of the binary expansion of
$\alpha$ is a zero for every $i \ge p$. The Lebesgue \me\ makes these events
independent and equally likely with probability $1/2$. Hence
(\ref{escape2}) can only occur for a null-\me\ set of $\alpha$'s.
This proves that for almost no $\alpha$'s an \eo\ can start from vertex
$V$ and pass through $j$ boxes before taking off to infinity. Since
events like this are countably many, we see that an \eo\ can almost never
leave from a vertex .
\qed
\skippar
Let us call $D_1$ and $D_2$ the sets of directions that satisfy, 
respectively, Corollary \ref{cor-main} and Lemma 
\ref{lemma-no-vertex}.  So $D:= D_1 \cap D_2$ is the ``full-\me'' set of 
directions that have all the generic properties we have analyzed so 
far.

\begin{corollary}
	For a.a. $\alpha$'s, the \bi\ flow $\fat$ over $R_{\alpha}$ around
	the \eo\ $\eta_{\alpha}$ is a local isometry. This means that, fixed
	a $z_{0} \in \eta_{\alpha}$, then $\forall T>0,\ \exists \eps >0$ s.t.
	\begin{displaymath}
		| z-z_{0} | \le \eps \Longrightarrow | \fat(z) - \fat(z_{0}) | = 
		| z-z_{0} | \ \ \ \forall t\in [-T/2,T/2].
	\end{displaymath}
	\label{cor-isometry}
\end{corollary}

{\sc Proof.} Since the \bi\ flow over $R_{\alpha}$ is isometric far
from the singular vertices, it suffices to observe that, for $\alpha
\in D$, $\eta_{\alpha}$ does not intersect any vertices. As regards 
the backward part, this is an immediate consequence of Lemma 
\ref{lemma-no-vertex} (since $\alpha\in D_{2}$).  The same holds for 
the escaping part, because a singular \eo\ would imply 
$n.i.(E_{\alpha}^{(n)}) = 2$ for $n$ large, and this cannot occur for 
$\alpha\in D_{1}$.
\qed
\skippar
We recall Leontovich's notation ``oscillating'', as introduced in
Section \ref{sec-exp}.

\begin{theorem}
	For almost all $\alpha$'s, the unique \eo\ is oscillating in the 
	past.
	\label{thm-backwards}
\end{theorem}

\begin{figure}[ht]
	\centerline{\psfig{file=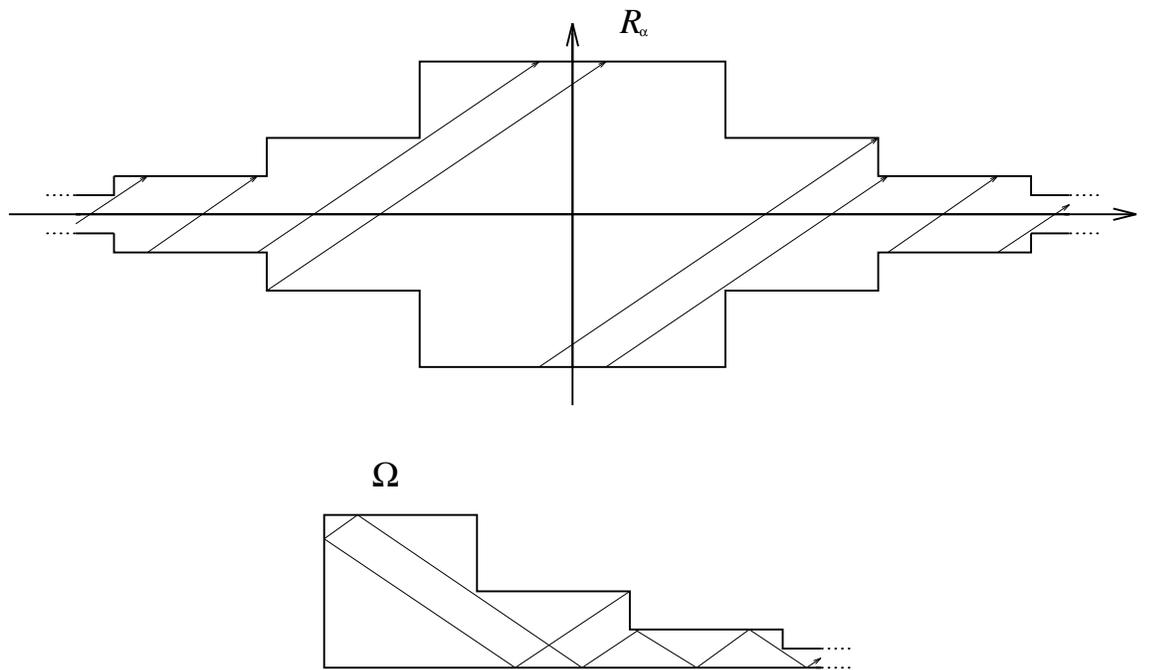}}
	\caption{A trajectory can be run over twice only if it contains a 
	non-singular vertex. For a.a. $\alpha$'s this is the only 
	possibility to have an orbit which escapes both in the past and 
	in the future.}
	\protect\label{fig8}
\end{figure}

{\sc Proof of Theorem \ref{thm-backwards}.} Let us take $\alpha\in D$,
again, and consider the unique \eo\ $\eta_{\alpha}$.  Lemma
\ref{lemma-no-vertex} states that it is non-singular, so the symmetry
arguments outlined in the remark in Section \ref{sec-exp} apply.
Suppose now that $\eta_{\alpha}^{-}$, some past semi-\tr y of
$\eta_{\alpha}$, escapes: this corresponds, by reflection, to a
forward escape semi-\o.  Then the uniqueness hypothesis shows that the
reflected image of $\eta_{\alpha}^{-}$ must coincide with some
$\eta_{\alpha}^{+}$.  In other words $\eta_{\alpha}$ is symmetric
around the origin in $R_{\alpha}$, which means that in $\Omega$ it is
run over twice, once for each direction.  The situation is
illustrated, for both $\Omega$ and $R_{\alpha}$, in Fig.  \ref{fig8}.
One gets easily convinced that the only way to realize this case is
that the \tr y has a point in which the velocity is inverted.  This
can only be a non-singular vertex.  But $\alpha\in D_2$ and Lemma
\ref{lemma-no-vertex} claims that this is impossible.
\qed

\section{Dynamics on the Billiard} \label{sec-dyn}

Throughout this section we fix a direction $\alpha\in D$.  As defined
in Section \ref{subs-back}, this is the set of directions satisfying all
the generic properties which we have explored so far.  Hence, for
simplicity, we drop the subscript $\alpha$ from all the notation.  For
example, the unique \eo\ will only be denoted by $\eta$.  On it, we
fix the {\em standard initial condition} $z_{0} = (0,Y_{0})$ as the last
intersection point with $L$, before the \o\ escapes towards $\infty$.
\par
Rather unexpectedly, it turns out that the statements in Section
\ref{subs-eo}, mainly intended to analyze the \eo s, provide, as a
by-product, a certain amount of information about the topology of the
flow $\ft$ on the \bi. Information which, although certainly
incomplete, we believe was not {\em a priori} obvious. The crucial
fact, as it will be noticed, is Corollary \ref{cor-main}, which roughly
states that not only there is just one initial point that takes
a \tr y to infinity, but also there is just one
neighborhood---necessarily around that point---that takes a
\tr y far enough. This is the idea behind next result.

\begin{lemma}
	Let $\alpha\in D$. Taken an \o\ $\gamma$, two numbers $\eps,T >0$,
	there exists a $w\in\gamma \cap L$, such that
	\begin{displaymath}
		| \ft(w) - \ft(z_{0}) | = | w-z_{0} | \le \eps \ \ \
		\forall t\in [-T/2, T/2] ,
	\end{displaymath}
	where $z_{0}$ is the standard initial condition on $\eta$.
	Furthermore, if $\gamma \ne \eta$, $w$ can be chosen arbitrarily far
	in the past or in the future of $\gamma$. For $\gamma=\eta$, $w$ can
	be chosen arbitrarily far in the past.
	\label{lemma-isom}
\end{lemma}

{\sc Proof.} Since $\alpha$ is typical, we can apply Corollary
\ref{cor-isometry} with $z_{0}$ fixed as above. This will return
an $\eps'$ (depending on $T$) such that all points as close to
$z_{0}$ as $\eps'$ remain such under the flow, within a time $T$.
Assume $\eps' \le \eps$ (if not, $\eps' := \eps$ will do). We need to
find a point of $\gamma$ in the interval $[Y_{0}-\eps',Y_{0}+\eps']
\subseteq L$. Recalling Corollary \ref{cor-main}, consider the
subsequence $\{ G_{n_{j}} \}$ of apertures whose backward beam of \tr
ies does not split at any vertex before reaching $L$.  Take a $j$ such
that $2 p_{n_{j}} = 2^{-n_{j}+1} \le \eps'$.  Since $\gamma$ is
unbounded, we can find a point $u \in \gamma \cap G_{n_{j}}$.  Call
$w$ the last intersection point of $\gamma$ with $L$, before $u$ is
reached.  From the non-splitting property of $G_{n_{j}}$, $|w - z_{0}|
\le \eps'$.  Corollary \ref{cor-isometry} shows that this is the sought $w$.
\par
Proposition \ref{prop-unbounded} actually states that {\em each}
semi-\tr y of $\gamma \ne \eta$ is oscillating: therefore $u$ (and so $w$)
can be chosen with as much freedom as claimed in the last statement of
the lemma.  As for $\eta$, only the backward part oscillates (Theorem
\ref{thm-backwards}).
\qed
\skippar
{\sc Remark.} We stress once again that the above is more than an
easy corollary of Proposition \ref{prop-truncated}: not only $\gamma$
and $\eta$ get close near infinity, being both squeezed inside the narrow
``cusp'', but, to be so, they must have already been as close for a
long time.
\skippar
A number of trivially checkable consequences of Lemma \ref{lemma-isom}
are listed in the sequel.  Recall the definitions of $\omega$-limit
and $\alpha$-limit of an \o\ as the sets of its accumulation points
in the future and in the past, respectively (see, e.g. \cite{w},
Definition 5.4).

\begin{corollary}
	With the same assumptions and notation as above,
	\begin{itemize}
		\item[(i)]  The \eo\ $\eta$ is contained in the
		$\omega$-limit and in the $\alpha$-limit of every other \o.
		\item[(ii)]  The \eo\ $\eta$ is contained in its own
		$\alpha$-limit.
		\item[(iii)]  Every invariant continuous function is
		constant.
		\item[(iv)]  The flow is minimal \iff the \eo\ is dense.
	\end{itemize}
	\label{cor-isom}
\end{corollary}

Of course, one would like to prove one definite topological property
of the flow $\ft$, such as minimality or at least topological
transitivity. Our techniques do not seem to do this job. However,
they do furnish a picture of how chaotic the motion on the \bi\ can
be. In fact, the attractor that $\eta$ has been proven to be is
certainly far from simple. Either it densely fills the whole
invariant surface $R_{\alpha}$, or it is a fractal set.

\begin{theorem}
	For a typical direction ($\alpha\in D$) consider the corresponding
	flow on $R_{\alpha}$. Denote $L_{\eta} := \eta \cap L$, the
	``trace'' of the \eo\ with the usual Poincar\'e section. Then its
	closure in $L$ (denoted by $\overline{L_{\gamma}}$) is either
	the entire $L$ or a Cantor set.
	\label{thm-attractor}
\end{theorem}

{\sc Proof of Theorem \ref{thm-attractor}.} Assume
$\overline{L_{\gamma}} \ne L$.  This set is closed.  We are going to
show it also has empty interior and no isolated points, that is, it is
Cantor.  In the remainder, by interval we will always mean a segment
of $L$.
\par
Suppose the interior of our set is not empty.  Then there exists an
open interval $I\subseteq \overline{L_{\gamma}}$ containing a point
$z$ of $\eta$.  Now, in the complementary set of
$\overline{L_{\gamma}}$, select a point $w$ whose \o\ is non-singular.
Let $w$ evolve, e.g., in the future.  By Corollary \ref{cor-isom},(i)
applied to $z$, there is a $t > 0$ such that $\phi^{t}(w) \in I$.  By
the choice of $w$, we can find an open interval $J$, such that $w\in
J$, $J \cap \overline{L_{\gamma}} = \emptyset$ and so small that
$\phi^{t}$ maps $J$ isometrically into $I$.  This implies that $J
\subset \overline{L_{\gamma}}$, which is a contradiction.
\par
To show that there are no isolated points: if $z\in
\overline{L_{\gamma}} \setminus L_{\gamma}$, there is nothing to
prove; if $z\in L_{\gamma}$, then Corollary \ref{cor-isom},(ii) does
the job.
\qed

\section{Conclusions}

Although we think we have given a pretty good description of the
\eo s for our model, the exponential step \bi, and we have concluded 
that those objects are central for the dynamics, the results contained 
in this paper certainly lack completeness.  Even conceding on 
semiclassical quantum mechanics, one is not satisfied from the point 
of view of ergodic theory, either.  Recalling Theorem 
\ref{thm-attractor}, we do believe that the flow should be minimal for 
a.e. $\alpha$, making the Cantor set case an interesting exception.  
But this does not seem to be easily provable with our techniques, which, we 
readily admit, use the results for finite polygonal \bi s (see 
Propositions \ref{mainRational} and \ref{mainBilliard}) blindly, 
without trying to extend them to our case. Most likely, doing so 
will provide a key to more complete statements.
\par
However, there is already something more to say on the \eo s for other 
models of infinite step \bi s.  Giving up the sharpness of the 
statements in Section \ref{subs-eo}, strictly designed for the case 
$p_{n}=2^{-n}$, a result similar to Theorem \ref{thm-generic} is 
at present available for a variety of cases. This is based on some 
elementary \me-theory and has the advantage that it does not require 
the {\em exact} knowledge of the behavior of the singular semi-\o s 
$\gamma_{p}$ as a function of $\alpha$ [as in 
(\ref{gammap-1})-(\ref{gammap-2})]. We refer the interested reader to 
\cite{ddl}.

\section{Acknowledgments}

The authors wish to thank S. De Bi\`evre, S. Graffi and Ya. G. Sinai
for stimulating discussions.  Also, M.L. would like to express his
gratitude to Ya. G. Sinai for his guidance during these years at
Princeton University. The same author has been partially supported by
the I.N.F.N. (Gruppo IV, Bologna).

\appendix
\section{Appendix: \\ The Binary Expansion of a Number} 
\label{app-expans}

Let $y$ be a real number. Suppose we want to analyze its binary
expansion: there exist an $m\in\Z$ such that
\begin{equation}
	y = \pm \sum_{j=m}^{-\infty} y^{(j)} 2^{-j} =: y^{(m)} \ldots
	y^{(0)}. y^{(-1)} y^{(-2)} \ldots;
	\label{bin-expans}
\end{equation}
$y^{(j)}=0$ or $1$. In order for this expansion to be in a one-to-one
correspondence with $\R$, we adopt the following convention: when
$y>0$ all endless sequences of the type $0111111\ldots$ are replaced by
$1000000\ldots$; if $y<0$ the rule is inverted. So, for instance,
$1/2 = 0.100\ldots$ and $-1/2 = -0.0111\ldots$
\par
The above convention ensures that numbers in $[-1,1[$ are described
with no need of integer digits, i.e. $y^{(j)} = 0,\, \forall j\ge 0$.
Recall that $y \mtwo$ means the unique real number in $[-1,1[$ congruent 
to $y$ modulo $2$.

\begin{lemma}
	Let $y\in\R$ and $\{ y^{(j)} \}_{j\le m}$ its binary expansion as 
	in {\rm (\ref{bin-expans})}.  Then $2^{k} y \in [-1/2,1/2[$ \iff 
	$y^{(-k-1)} = 0$.  \label{lemma-expans}
\end{lemma}

The trivial proof is omitted.

\vfill\eject

\noindent{\Large\bf Figures}
\medskip

\medskip
\medskip
\medskip
\medskip
\medskip
\medskip
\medskip

\end{document}